# A group-IV double heterostructure light emitting diode for room temperature gain in Silicon

Andreas Salomon[1], Johannes Aberl[1], Lada Vukušić[1], Enrique Prado-Navarrete[1], Jacqueline Marböck[1], Diego-Haya Enriquez[1], Jeffrey Schuster[1], Kari Martinez[2], Heiko Groiss[2], Thomas Fromherz[1,*], and Moritz Brehm[1,*]

[1] *Institute of Semiconductor and Solid State Physics, Johannes Kepler University Linz, Altenberger Strasse 69, A-4040 Linz, Austria*

[2] *Christian Doppler Laboratory for Nanoscale Phase Transformations, Center for Surface and Nanoanalytics, Johannes Kepler University Linz, Altenberger Straße 69, 4040 Linz, Austria*

* Corresponding authors: *thomas.fromherz@jku.at ; moritz.brehm@jku.at*

## Abstract

**The lack of straightforward epitaxial integration of useful telecom lasers on silicon remains the major bottleneck for bringing optical interconnect technology down to the on-chip level. Crystalline silicon itself, an indirect semiconductor, is a poor light emitter. Here, we identify conceptionally simple $Si/Si_{1-x}Ge_x/Si$ double heterostructures (DHS) with large Ge content ($x \gtrsim 0.4$) as auspicious gain material suitable for Si-based integrated optics. In particular, using self-consistent Poisson-current transport calculations, we show that Si diodes containing a 16 nm thick $Si_{1-x}Ge_x$ layer of high crystalline quality, centered at the p-n junction, results in efficient carrier accumulation in the DHS and gain if the diode is driven in forward direction. Despite the high strain, we unambiguously demonstrate that such prior unattainable defect-free DHS can be fabricated using ultra-low temperature epitaxy at pristine growth pressures. Telecom light emission is persistent up to 360 K, and directly linked to a ~160 meV high conduction band barrier for minority electron injection. This epitaxy approach allows further increasing the Ge content in the DHS and creating dot-in-well heterostructures for which even higher gains are predicted. Thus, the surprisingly facile DHS presented here can be an essential step toward novel classes of group-IV optoelectronic devices for silicon photonics.**

Double heterostructure (DHS) diodes for optoelectronic applications offer vast benefits as compared to conventional semiconductor diodes through charge carrier accumulation at the p-n junction, thereby favoring recombination over diffusion currents, tuning possibility of the emission wavelength, population inversion, and refractive index guiding of light.[1-3] The suppression of diffusion currents by the DHS concept results in order of magnitudes reduced threshold currents in electrically-pumped diode lasers, and was the enabling technology for the widespread application of semiconductor lasers realized in the strain-free AlGaAs/GaAs epitaxial system. Nowadays, great efforts are being undertaken to bring optoelectronic functionality to the Si platform through Si photonics.[4-11] However, to-date, no approach combines the entire Si photonic emitter wishlist, including electrical pumping, emission at telecom wavelengths, above room-temperature operation, direct (epitaxial) and parallel integration, convenient and efficient coupling to Si waveguides and modulators, output efficiency, thermal budget, low costs, and compatibility with Si electronics.[12]

For group IV materials, DHS have been realized only for highly-diluted SiGe alloys and applications in electronics.[13] A practical use of the DHS concept in light-emitting diodes (LEDs) operating above room temperature (RT), requires higher Ge concentrations, $c_{Ge}$. However, the significant lattice mismatch between Si and SiGe increases approximately linearly with $c_{Ge}$, limiting the thickness for defect-free layers grown on Si. Above a critical thickness, strained layers relax elastically via surface roughening or plastically through forming stacking faults, misfit dislocations or/and point defects. Therefore, conventional wisdom dating back to the 1960s[14,15] suggested the need of high epitaxial growth temperatures ($T_G$) of >500°C to keep point defect densities low and the epitaxial quality high.[16] However, at these common $T_G$s DHS LED structures suitable for RT operation cannot be formed since these require high $c_{Ge} \geq 0.4$ to induce large energy band offsets, combined with a layer thickness above ~10 nm to minimize quantum confinement that, in turn, would counteract the benefits of high band offsets. Additionally, concerning laser applications, a wider SiGe layer enhances the overlap of an optical laser mode with the gain, potentially provided by the SiGe layer.

Here we show, based on self-consistent Poisson-current transport calculations, that already a simple DHS structure, 16 nm thick of $Si_{1-x}Ge_x$ with $x \geq 0.4$ sandwiched between p- and n-doped Si layers with the p-n junction at its center, leads to efficient carrier accumulation of both electrons and holes in the SiGe layer, upon applying a forward bias similar to $E_{SiGe}/e$, where $E_{SiGe}$ denotes the bandgap energy of the SiGe layer. In this case, at RT, significant carrier inversion builds up in the SiGe layer and gain in the range between 50-300 $cm^{-1}$ is calculated at telecom wavelengths for forward current densities between 8-30 kA/cm$^2$, where the higher gain values were obtained for $c_{Ge}$ = 60%. Further, we unambiguously demonstrate that such DHS layers can be realized with outstanding crystalline quality using epitaxy at unconventionally low $T_G$ and pristine maximum chamber pressures < $2\cdot10^{-10}$ mbar during growth.

Fabricated LEDs exhibit pronounced and temperature-persistent electroluminescence operating well above RT, unusual for group-IV light emitters. These findings demonstrate the vast potential for this emerging gain material that can be readily integrated into Si platforms for telecom applications using Si technology.

## Results

### Theoretical assessment of SiGe DHS with high Ge content

Figure 1 indicates the dramatic combined effects of p- and n-doping, and forward biasing on the band edge profiles of a 16 nm thick $Si_{0.6}Ge_{0.4}$ layer sandwiched between two Si cladding layers. For such a DHS, the profiles of the conduction band (CB) minima at the six Δ-points of the Brillouin zone across the DHS are shown by the orange and blue lines in Fig. 1a,b. For the valence band, the profiles for the two lowest hole energy bands [heavy hole, (HH), light hole (LH)] with minima at the Γ point are shown by green and red lines in Fig. 1a,b, respectively. Figure 1a clearly shows, that in the undoped DHS, holes (h) can minimize their energy by accumulating in the SiGe layer, whereas for electrons (e) the SiGe layer is energetically not favorable. Thus, electrons are not attracted to it. Such a band ordering across an interface is called type-II alignment and has to be clearly distinguished from type-I alignment present, for example, in AlGaAs/GaAs/AlGaAs DHS, where *the GaAs layer attracts both e and h*.

Note that due to the larger lattice constant of bulk SiGe as compared to Si, SiGe is biaxial compressively strained perpendicular to the direction of the layer sequence if, during the epitaxial growth at sufficiently low $T_G$s, no misfit dislocations were incorporated. As a consequence of this strain, both the six-fold and two-fold degeneracies of the conduction and valence minima, respectively, are lifted. Thus, the conduction band minima with the lowest energies become those four with an electron wavevector pointing perpendicular to the layer sequence direction (orange line, labeled $\Delta_{xy}$ in Fig. 1a,b). The other two conduction band minima oriented along the growth direction (blue line, labeled $\Delta_z$ in Fig. 1a,b) are so much raised in energy above the $\Delta_{xy}$ ones that they become irrelevant for transport. The HH band extremum at the Γ-point becomes the ground state for holes.

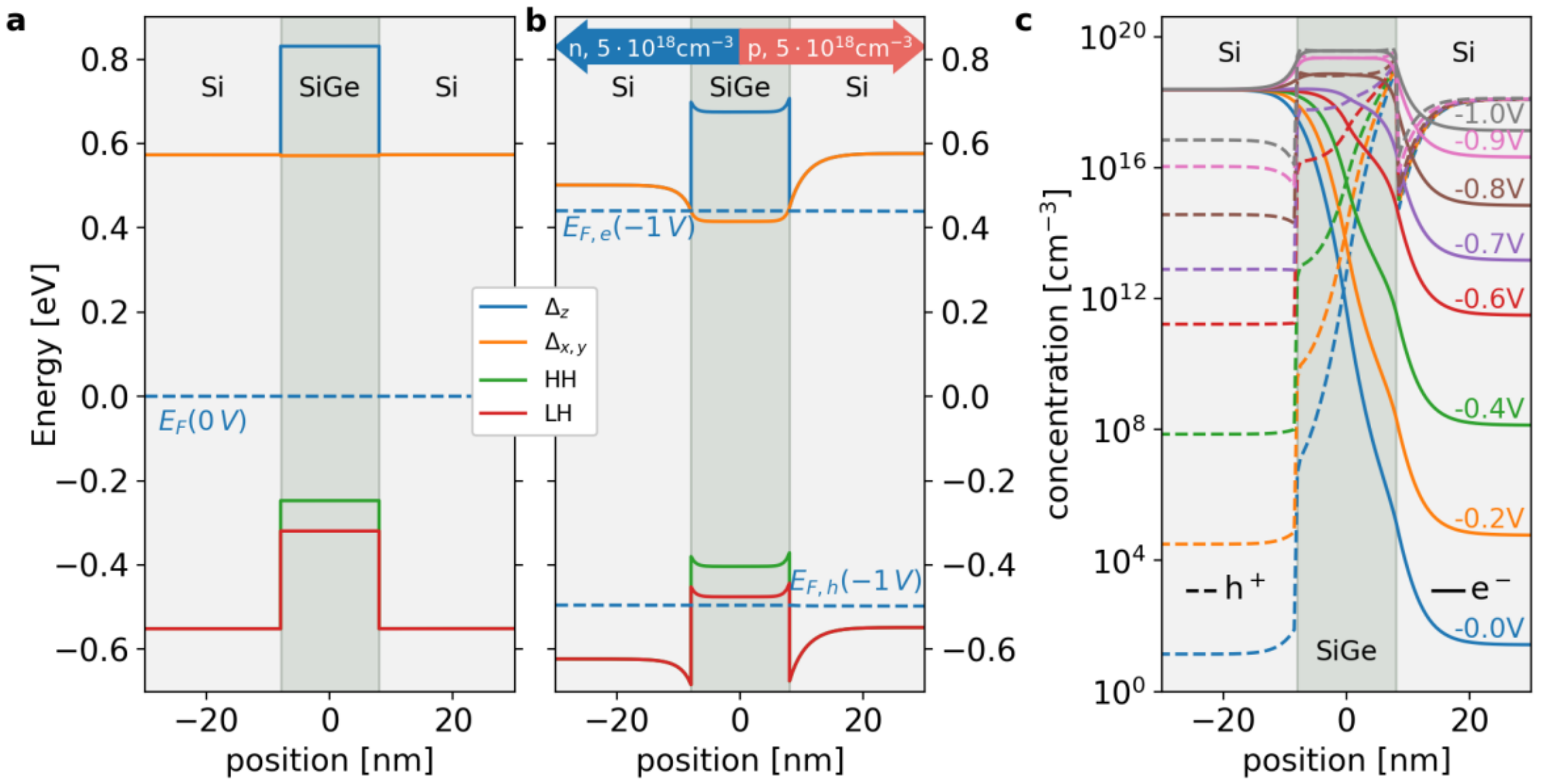

**Fig. 1| Band edge- and charge carrier concentration profiles across a $Si/Si_{0.6}Ge_{0.4}/Si$ double heterostructure with a 16 nm thick alloy layer. a** Undoped DHS; **b** DHS with centered p-n junction, biased into forward direction to approximately compensate for the built-in voltage. The profiles of the conduction band minima at the Δ-points of the Brillouin zone oriented laterally and parallel to the growth direction ($\Delta_{xy}$, $\Delta_z$) are shown by the orange and blue lines, respectively, the profiles of the heavy (HH) and light (LH) hole valence bands at the Γ point by the green and red lines**.** The doping regions, types and concentrations are indicated by the arrows. Due to the external bias, the Fermi energies for electrons ($E_{F,e}$) and holes ($E_{F,h}$) are split; **c** Electron (full lines) and hole (broken lines) concentration profiles at the various forward bias values are given as labels**.** The external bias is measured with respect to the contact to the p-doped side of the DHS kept at bias ground.

P- and n-type doping the Si and SiGe layers to a concentration of $10^{18}$-$10^{19}$ cm$^{-3}$ with the p-n junction approximately in the center of the SiGe layer and compensating the resulting built-in voltage of ~1V by an externally applied bias voltage $V_B$ radically changes the alignment of the conduction-band edges in Si and SiGe as shown in Fig. 1b. The shown band edge profiles result from strain-dependent Poisson-current transport calculations.[17] Shockley-Read-Hall, radiative, and Auger recombination processes were included in these simulations. Details on the respective recombination constants are given in the Supplementary Material.

Similar to a type-I alignment, in this situation, the $\Delta_{xy}$ conduction band minima in the SiGe layer become the lowest conduction band states. Thus, in addition to the holes in the HH valence band, also the electrons experience an attractive potential by the SiGe layer, and spatially overlapping e and h-distributions within the SiGe layer are established, dramatically enhancing the radiative e–h recombination rate as discussed below.

Figure 1c shows the simulated effect of the external bias voltage on the local carrier concentration near the DHS, where the bias values are given in the figure legend. In the n-type region, a constant donor concentration of $5 \cdot 10^{18}cm^{-3}$ was assumed, whereas in the p-region, for the 58 nm next to the p-n junction and for the remaining 150 nm acceptor concentrations of $5 \cdot 10^{18}cm^{-3}$ and $2 \cdot 10^{19}cm^{-3}$, respectively, were used in the simulations. Figure 1c clearly shows that for bias voltages more negative than ~ -0.8V, both the h (dashed lines) and e concentrations (solid lines) in the SiGe layer become larger than the respective doping concentrations in the Si cladding regions, both reaching $3.7 \cdot 10^{19}cm^{-3}$ at -1V bias. For such bias voltage levels, the carrier concentration increase per voltage step becomes less effective, as can be clearly seen in Fig. 1c by comparing the carrier concentration profiles for the bias voltages -0.8V to -1V. The reason for this efficiency drop is the exponential increase of the diode's forward e and h diffusion currents injected into the Si cladding layers, effectively counteracting the carrier accumulation in the SiGe layer.

This efficiency decrease in the carrier accumulation with increasing forward bias, and thus, with increasing diode current density is more clearly displayed in the inset of Fig. 2, which shows the ratio between the simulated current due to radiative e-h recombination within the 16 nm thick SiGe layer and the total diode current. Since outside this layer, the radiative recombination is negligible, this ratio corresponds to the internal quantum efficiency (IQE) and shows a clear maximum around 0.5 $A/cm^2$ total current density for the $Si_{1-x}Ge_x$ DHS diode with $x \geq 0.4$ at 300 K. By increasing $c_{Ge}$, the minority carrier injection into the Si cladding layers, and, thus, the minority carrier diffusion currents, become increasingly suppressed by the CB and VB barriers between Si and SiGe. Most importantly, as compared to the simulation results obtained for the radiative recombination within ± 8nm centered around the p-n junction of a Si-only diode (violet line), at 300 K, the $Si/Si_{0.6}Ge_{0.4}/Si$ DHS diode is predicted to show more than four orders of magnitude increased IQE. Figure 2 shows that the DHS - enhancement of the radiative recombination rate saturates at T = 300 K around a Ge concentration of 50-60%. At these concentrations, the e- and h- diffusion currents from the SiGe layer into the Si regions are already so much suppressed by the potential barriers, that not much is gained by further increasing them.

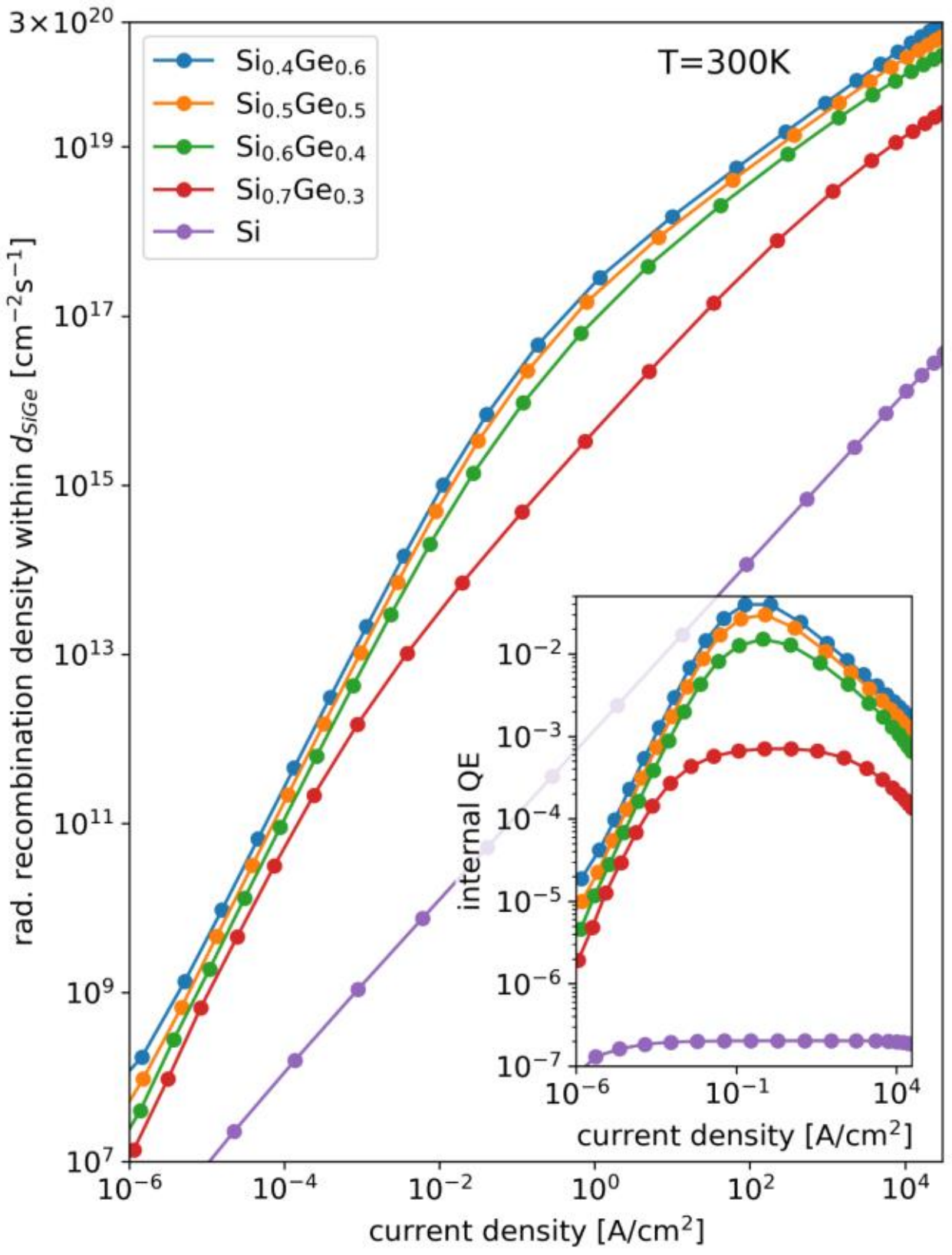

**Fig. 2| Radiative recombination and internal quantum efficiency at room temperature.** Comparison of simulated areal recombination densities as functions of the injected current densities within the SiGe layer of $Si/Si_{1-x}Ge_x/Si$ DHS p-n diodes for four $c_{Ge}$ values $0.3 \leq x \leq 0.6$. Radiative recombination events outside the 16 nm thick SiGe layers contribute insignificantly to the respective curves. For comparison, for a Si-only diode with the same Si layer thicknesses and doping levels the radiative recombination density within +/- 8nm from the pn junction are shown in violet. For the same structures, the inset shows the simulated internal quantum efficiency for radiative recombination calculated as ratio between radiative recombination- and total current-densities.

The e-h recombination parameters used in this work are listed in Tab. 3 of the methods section. While the simulated numbers for the IQE depend, of course, strongly on these values, the *ratio* of the results for the DHS and the Si-only diode, which are based on the same set of linearly interpolated values for Si and Ge with respect to the Ge concentration, can be expected to be much less sensitive to these parameter values. Thus, Fig. 2 demonstrates the enormous improvement resulting from the DHS concept for type-II band alignment systems like the Si-Ge material. The main plot of Fig. 2 compares the areal spontaneous radiative recombination rate densities as a function of the diode current simulated for several Ge fractions in $Si/Si_{1-x}Ge_x/Si$ DHS at 300 K and for the corresponding plain Si diode. The plotted simulation results are thus proportional to what is observed in typical L-I experiments. Also, in this plot, at 300 K at least 3-4 orders of magnitude of spontaneous emission enhancement due to the insertion of only 16 nm $Si_{1-x}Ge_x$ with $x \geq 0.4$ into a Si diode to form a DHS are predicted over the experimentally relevant current density region between $10^{-2}$ - $3 \cdot 10^4$ A/cm$^2$.

**Gain in group-IV double heterostructures**

Figure 1b shows that biasing the DHS diode by -1V in forward direction results in electron and hole quasi-Fermi energies ($E_{F,e}$ and $E_{F,h}$) located in the conduction and valence band, respectively. In such a situation, optical gain can be expected for photon energies $E_{\Delta xy}$-$E_{HH}$ < $\hbar\omega$ < $E_{F,e}$ - $E_{F,h}$. Indeed, the simulated gain spectra for a $Si/Si_{0.4}Ge_{0.6}/Si$ DHS diode displayed in Fig. 3a for absolute forward bias values in the range 0.55 – 1.2V, clearly show positive values for absolute bias larger than 0.75V. On the left ordinate axis of Fig. 3a, the gain coefficient of the SiGe layer is plotted vs. the photon energy on the lower abscissa axis, whereas the simulated I-V characteristics [blue, monotonic line in Fig. 3a] is plotted using the blue right ordinate and top abscissa axes. The various colors of the gain spectra correspond to the colors of the dots superimposed on the I-V curve, which indicate the respective bias conditions at which the gain spectra were simulated. As shown in Fig. 3a, the gain coefficient and its peak photon energy increase with increasing current density j.

At -1.2V forward bias (j = 30 kA/cm$^2$), a maximum gain coefficient of ~300 cm$^{-1}$ is calculated for a photon energy of 0.82 eV (1512 nm wavelength). Higher gain coefficients can be expected for even higher forward current densities, however, because of numerical issues, simulations are not reliable at such large current densities. A summary of the simulated dependence of the maximum gain of a 16 nm thick alloy layer in a $Si/Si_{1-x}Ge_x/Si$ DHS diode on the forward current density for various Ge fractions (x=0.3, 0.4, 0.5, 0.6) at 200 K and 300 K is shown in Fig. 3c, whereas Fig. 3b shows the photon energies at which these maxima occur in dependence of the bias current. The shaded energy region in Fig. 3b indicates the telecom wavelength region. As summarized by Fig. 3, our simulations thus indicate significant room temperature gain in the telecom spectral range in an only 16 nm thick alloy layer of Si/SiGe/Si DHS diodes with various compositions at moderate current densities.

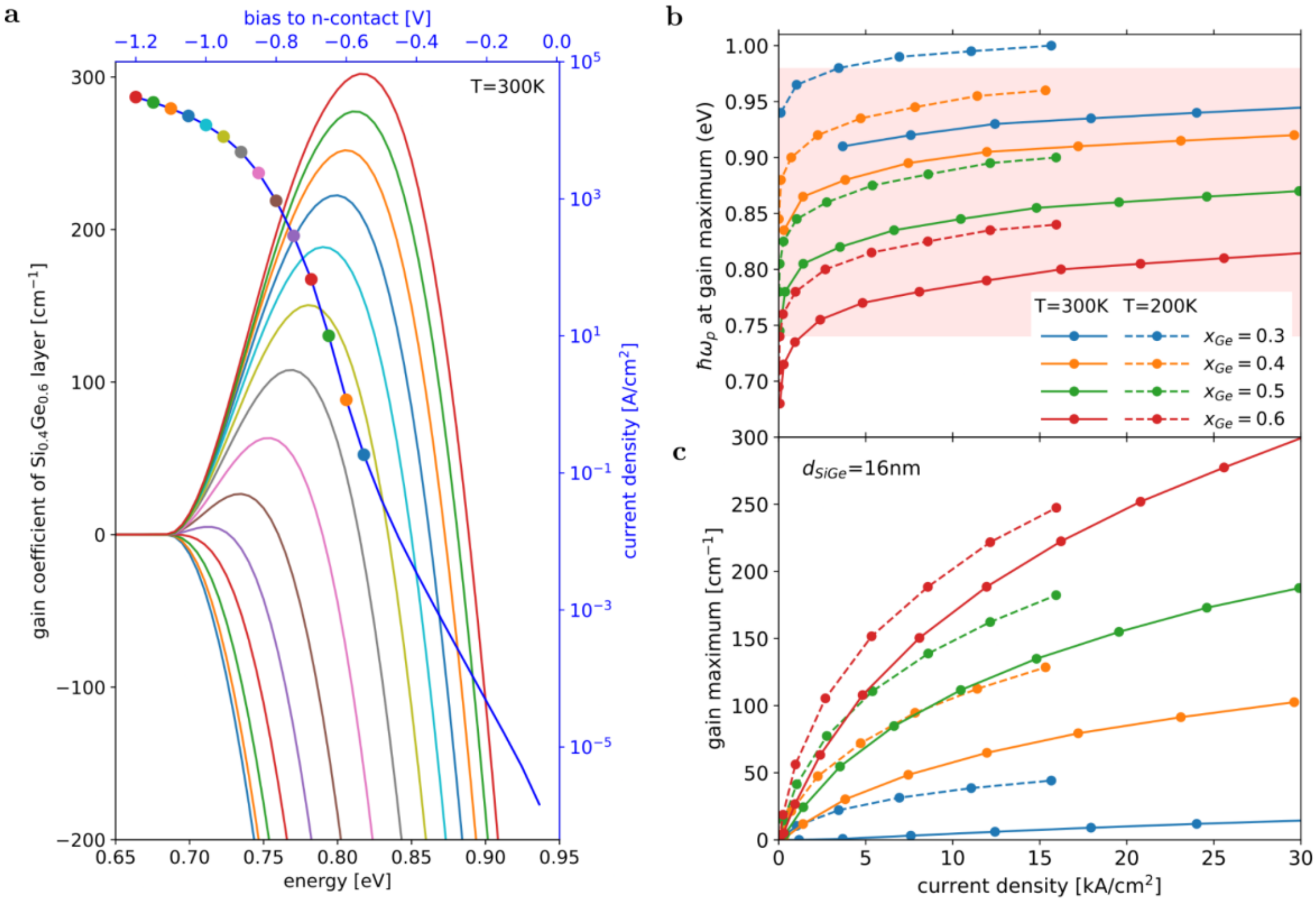

**Fig. 3| Gain in group-IV double heterostructures. a** Simulated room temperature gain spectra of the SiGe layer in a $Si/Si_{0.4}Ge_{0.6}/Si$ DHS diode at various bias conditions (black left- and bottom axes). The color of the gain spectra lines corresponds to the color of the dots superimposed on the simulated I-V characteristics of the DHS diode shown by the blue, monotonic line. These dots indicate on the blue plot axes (top and right) the bias- and current-pairs, for which the respective gain spectrum was simulated. **b** Shift of the photon energy $\hbar\omega_p$ at which the gain maximum occurs with the current density injected into a $Si/Si_{1-x}Ge_x/Si$ DHS diode for various alloy fractions x given in the legend for room temperature and 200K. The shading indicates the telecom wavelength region; **c**: Simulated gain coefficient of a 16 nm thick alloy layer in a $Si/Si_{1-x}Ge_x/Si$ DHS diode in dependence of the current density at 200 K and room temperature. The color- and line type-coding for the Ge fractions x and the temperature is given by the legend of **b**.

## Experimental implementation of group-IV DHS LEDs

Experimentally, DHS diode structures with structural parameters as used in the simulations (see transmission electron microscopy image in Fig. 4a) were grown using ultra-low temperature epitaxy and excellent maximum *growth* pressures $< 2\cdot10^{-10}$ mbar. Square mesa diodes with 400 µm sides were fabricated using standard Si processing techniques in a cleanroom environment. To enable electroluminescence emission perpendicular to the diode surface, the electrical metal (Au) contact to the top p-type layer was implemented as ring contact of 20 µm metal width along the circumference of the mesa. A typical temperature-dependent diode behaviour in the current-voltage (I-V) characteristics is observed, see Fig. 4b and inset of Fig. 4c. The first electroluminescence spectra from a group-IV DHS diode as observed in this work are depicted in Figs. 4c,d for the selected sample temperatures of 10 K and 295 K, respectively. At 10 K and excitation currents above 12 mA, we observe bulk-like light emission[18] at the expected wavelength, considering the SiGe alloy concentration and the strain in the system.[19]

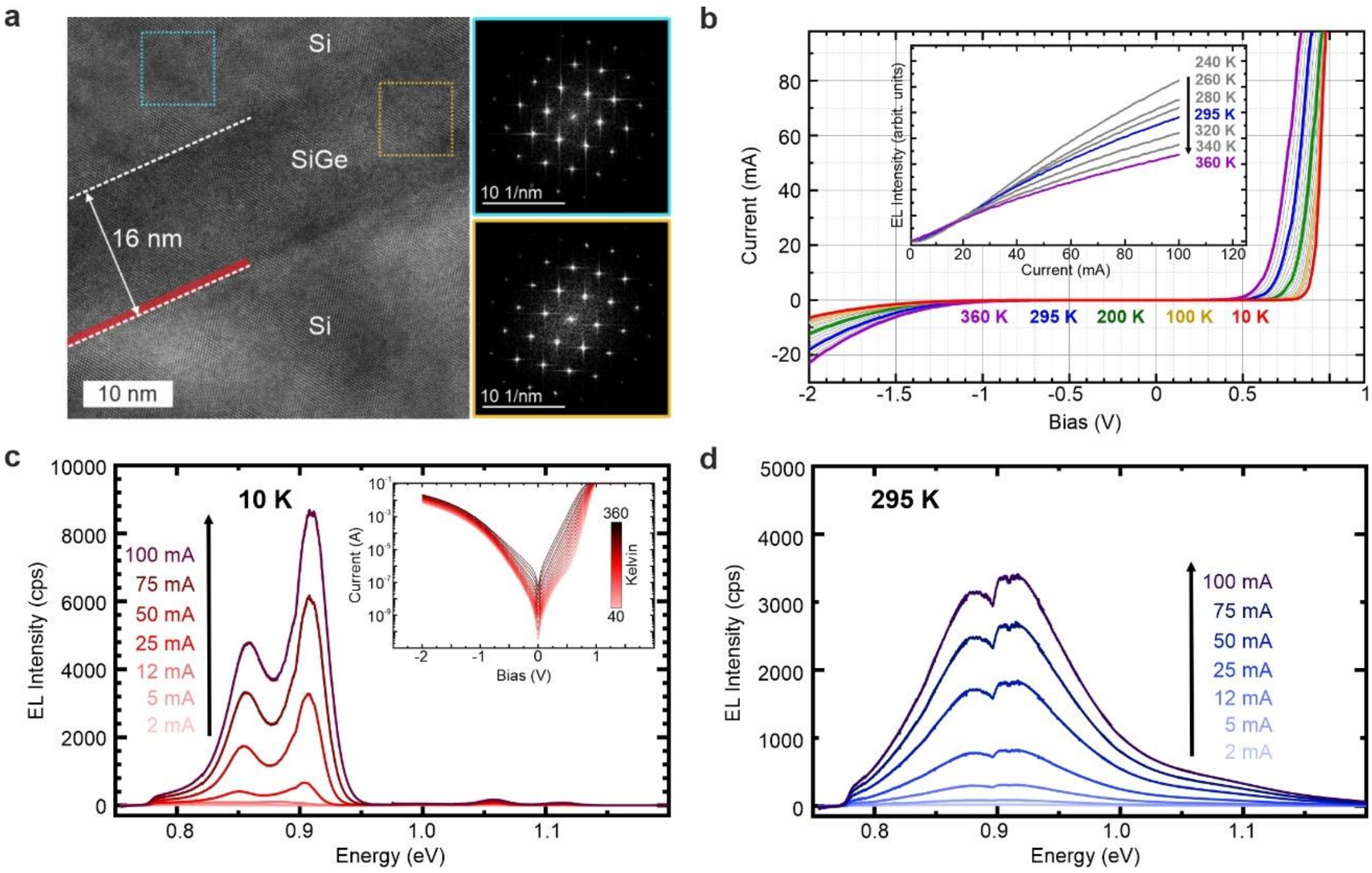

**Fig. | 4 Structural, electrical, and optoelectronic characterization of the DHS structure. a** Cross-sectional transmission electron microscopy image of the DHS, 16 nm $Si_{0.6}Ge_{0.4}$ sandwiched between crystalline Si layers. The Fast Fourier Transform patterns from the colour coded rectangles of the DHS and the Si matrix are shown on the right. The red-shaded area marks the maximum possible SiGe layer thickness using conventional high-temperature growth.[20] **b** Current-voltage characteristics for different temperatures. Inset: light-current curves above 240 K, indicating thermal quenching, but light emission up to 360 K; **c** Injection current-dependent electroluminescence spectra at 10 K and **d** at room temperature (295 K). The inset in **c** indicates the I-V characteristics of the diode on a logarithmic scale. The dips in the EL spectra at 0.897 eV are associated with absorption due to the fused Silica cryostat window.

In contrast to PL spectra from undoped DHS samples (see Ref. 20), for which type-II band alignment between the pseudomorphic SiGe and the Si substrate (see Fig. 1a) leads to a pronounced e-h wavefunction separation, the EL emission is robust and persists up to room temperature and above, atypical for group-IV light emitters. Based on our simulation results, we argue that at sufficiently large bias voltages (see Fig. 1b,c), the enhanced emission occurs due to the efficient confinement of electrons and holes in the DHS region. This confinement leads - similar to the prototypical AlGaAs/GaAs/AlGaAs type-I band alignment system - to an enhanced e-h wave function overlap, and strongly increased e and h concentrations in the diode's SiGe region (Fig. 1c). At 295 K, we observe an expected temperature-induced broadening of the EL emission, while the integrated intensity at high driving currents (100 mA) is as large as the one observed at 10K and only reduced by 23% compared to the maximum one observed at 240 K.

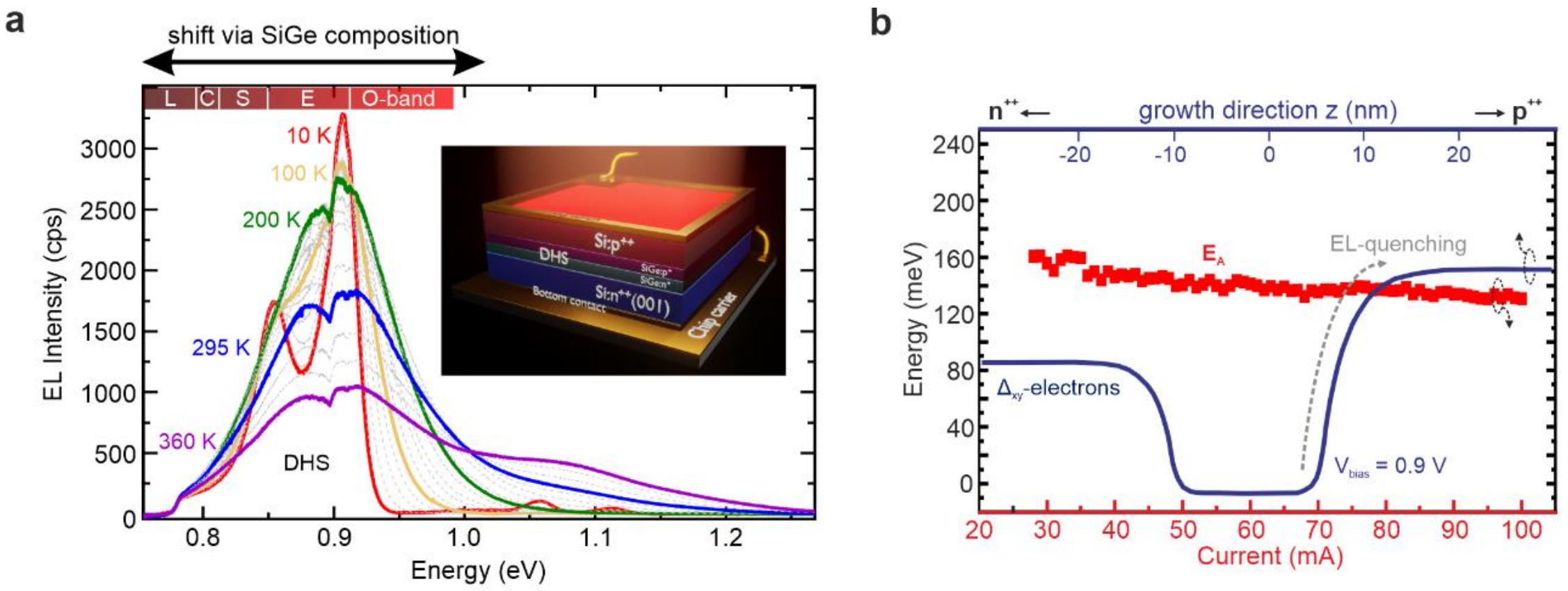

**Fig. 5| Temperature-dependence of the EL emission and activation energies for thermal emission quenching. a** EL spectra, recorded for an injection current of 50 mA and a sample temperature ranging from 10 K (red) to 360 K (violet). Grey dashed lines indicate the evolution of the spectra in temperature steps of 20 K. The inset schematically shows the diode structure. **b** red data points: Activation energies $E_A$ for thermal quenching vs. the driving current. Blue: $\Delta_{xy}$ conduction band minimum profile (at 300 K) for a forward bias of -0.9 V, indicating the agreement between the extracted $E_A$'s and the potential barrier height for electron minority carrier injection into the p doped region of the diode.

Finally, Fig. 5 shows the temperature-dependent spectra of the DHS diode at a driving current of 50 mA, indicating pronounced emission up to 360 K (87°C), the temperature limit of our setup. From the current and temperature-dependent EL emission data, we systematically extracted the activation energies for thermal EL quenching from Arrhenius plots according to $I(T) = I_0 \cdot \left(1 + A \cdot exp\left(\frac{-E_A}{k_B T}\right)\right)^{-1}$, where I(T) is the temperature-dependent integrated intensity, $I_0$ is the intensity at 0 K, $E_A$ is the activation energy, $k_B$ is the Boltzmann constant, T is the temperature, and A is a fitting parameter. The current-dependent activation energies $E_A$ are depicted in Fig. 5b, overlayed with the CB edge profile of the DHS-diode at 300 K and at a forward bias voltage of -0.9 V. In a current injection range from 30 mA to 100 mA, $E_A$ is monotonously decreasing from ~160 meV to ~130 meV. This behaviour is in remarkable agreement with the height of the potential barrier forming across the DHS structure according to our simulations at bias voltages compensating the built-in voltage (see. Fig. 1b), which effectively limits electron minority carrier injection. We want to emphasize that the formation of this barrier is essential for this device, and its experimental verification is an important step toward the design and implementation of the more elaborate electrically pumped DHS laser devices based on the simulations presented in this work.

**Discussion**

The direct implementation of SiGe heterostructures on Si has been widely studied in the past decades to boost the performance characteristics of Si-based electronics and optoelectronics. As a common sense result, it has been thought that gain cannot be achieved from simple SiGe/Si(001) heterostructures, and the search for light sources that can be implemented on Si was widened to more exotic materials and material fabrication and implementation techniques. While for these exotic approaches - exotic in terms of the semiconductor industry - impressive progress has been reported, they come at the cost of a simple and straightforward implementation on Si(001) as they often include the need for micrometer-thick defective virtual substrates[7,10,21-24], Ge on insulator[25], non-standard crystal orientations[9,26,27] substrates, and/or materials that are still foreign to the semiconductor industry.

Here, however, we demonstrate a most simple and robust way to achieve room temperature gain in Si from a nanostructural geometry point of view. It comprises thin, fully strained, two-dimensional SiGe layers that can be directly grown on Si(001) surfaces used in the semiconductor industry. Our approach does not involve any complicated nanostructural geometries, interfaces, chemical compounds, or atomic configurations. Only flat layers of SiGe of high crystalline quality are sandwiched between doped Si crystals. Therefore, the theoretical validation of the here presented results is highly robust and can be calculated using simple one-dimensional transport models. We demonstrate that for SiGe DHS layers with a Ge concentration of 40% or higher, significant charge carrier inversion and gain can be achieved even at room temperature. Notably, the SiGe layer thickness has to be large enough to avoid substantial quantum confinement in the well that would shift the ground state above the Fermi levels, thus inhibiting gain. The room-temperature gain increases with increasing Ge concentration (Fig. 3c) to values up to 300 $cm^{-1}$.

Here, we also demonstrate that such DHS layer structures can be fabricated by the combination of ultra-low growth temperatures and excellent background pressure during the growth. At conventional growth temperatures <500°C, such DHS layers would elastically or plastically relax. In Fig. S6 of the supplementary material, we show that even at high Ge concentration $x = 0.63$, still flat and non-dislocated SiGe layers of 23 nm thickness can be grown.

Even though ultra-low temperatures are used during growth, we emphasize that these structures are very robust against thermal annealing. In the supplementary material, we demonstrate that the DHS layers can be thermally annealed at 500°C without significant change in the surface roughness or strain status. This finding points to the fact that ultra-low temperatures only need to be employed during the growth, and strain relaxation mechanisms that typically limit the layer thickness are dominantly linked to the surface kinetics during growth. The robust thermal budget of the ULT-SiGe DHS layers, in turn, allows for enhancing the complexity of the DHS. This can be done, e.g., by creating dot-in-well structures using hut clusters[28] (see Fig. S2,S3,S4), or embedding gain material like color centers[29] through their self-assembly,[30,31] or quantum dots, for which their light emission is boosted by defects[32,33] into the DHS layer. Since our approach can be directly implemented on Si, standard SOI technology can be used to use these novel DHS to form photonic resonators, as electrically-driven microdisk lasers and ridge lasers can be envisioned.

Also, by lowering the growth temperature further, higher layer supersaturation and, thus, even thicker dislocation-free layer thickness at a given Ge concentration can be expected, see Fig. S6.

In conclusion, we have investigated the light-emission behavior of proof-of-principle group-IV double heterostructure diodes grown at ultra-low temperatures. Experiments and transport bandstructure calculations indicate that high energy activation barriers for electron and hole minority carrier injection and type-I-like band edge profiles alignment under forward bias lead to pronounced light emission above room temperature. Already in this simple model system, room-temperature gain is predicted as the applied bias leads to charge carrier inversion within the lower bandgap SiGe layers. We note that these barriers could be enlarged without increasing the system’s strain by using graded SiGe layers with an *average* Ge content of 40% and the same layer SiGe layer thickness. Additionally, no particular effort has been undertaken to optimize contacts and sidewall passivation of the diodes so far. In such a straightforward system, the transition energies can be tuned by the SiGe alloy's Ge content, while the structures' width needs to stay large enough to avoid quantum confinement in the DHS. Despite the need for ULT growth, the thermal budget during DHS formation can be increased to at least 500°C, indicating that additional emitters, such as QDs, can be embedded in the DHS structure.

# Methods:

### Self-consistent Poisson-current transport calculations of double heterostructure diodes

Simulations of the bandedge profiles across the $Si/Si_{1-x}Ge_x/Si$ DHS diode were performed using the commercially available nextnano$^{++}$ program package.[17] The simulated effect of the external bias voltage on the local carrier concentration in the vicinity of the DHS, where the bias values are given in the legend of Fig. 1c, refer to the potential applied to contact of the Si n-type region at 5300 nm distance from the p-n junction relative to the grounded Si p-contact at 208 nm distance.

Since in the temperature region of interest (~RT), the confinement energies for electrons and holes in the SiGe layer are smaller than the thermal energy, electron and hole concentrations in all regions of the DHS were described classically. For the SiGe layer pseudomorphically strained to the Si substrate, the strain-induced splittings of the conduction and valence bands and the corresponding modifications in the density of states were included in the simulations. The corresponding elastic constants and deformation potential parameters used in the simulations are listed in Tab. 1.

Table 1: Elastic constants, deformation potentials, valence band spin-orbit splitting and Varshni parameters for Si and Ge used in this work. For the $Si_{1-x}Ge_x$ alloy, the corresponding values were obtained by linear interpolation with respect to x.

| Name | Symbol | Si | Ge |
|---|---|---|---|
| elastic constants [GP] | c11 | 165.77[34] | 128.53[34] |
| | c12 | 63.93[34] | 48.26[34] |
| absolute deformation potential for *Δ* conduction band [eV] | $a_c^{ind}$ | 3.4[35] | 0.14[35] |
| valence band deformation potential for shear strain of tetragonal symmetry [eV] | b | -2.10[36] | -2.86[37] |
| valence band spin-orbit splitting [eV] | $\Delta_{so}$ | 0.044[34] | 0.30[34] |
| *Δ* conduction band deformation potential for uniaxial strain along [001] [eV] | $\Xi_u^{\Delta}$ | 9.16[38] | 9.42[38] |
| Varshni parameter [$10^{-4}$eV/K] | α | 4.730[39] | 4.774[39] |
| Varshni parameter [K] | β | 636[39] | 235[39] |

For the dependence of the fundamental bandgap $E_g^{SiGe}$ of unstrained $Si_{1-x}Ge_x$ on the Ge fraction x at T = 4.2 K $E_{g,0}^{SiGe}(x) = 1.155 - 0.43x + 0.206x^2$ was used,[18] whereas its variation with the temperature was described by[40] $E_g^{SiGe}(T) = E_{g,0}^{SiGe}(x) - \alpha(x)T^2/[T+\beta(x)]$, with the Varshni parameters α(x) and β(x) linearly interpolated between the values given in Tab. 1. Note that the α- and β-values listed in Tab.1 for Ge refer to the Γ-L bandgap of Ge. Due to a lack of experimental data on the temperature dependence of the Γ-$\Delta$ bandgap, we estimated the corresponding α- and β-values by those listed in Tab. 1 for the Γ-L bandgap. The alignment of the average valence band energies of unstrained $Si_{1-x}Ge_x$ and Si was described by $\Delta V_{av}$=580[meV] · x, i.e., by the default relation included in the nextnano++ database. In literature, slightly different values for the prefactor or the Ge content were reported, ranging from 470 meV[41] - 540 meV.[38] Also, for the parameter values listed in Tab. 1, different values can be found in the literature. However, we decided also for these parameters to use the default values as defined in the nextnano++ database for two reasons: first, this choice facilitates the comparison with results of other nextnano++ users relying on the same database; second, since the simulations of this work are focused on qualitative trends with varying Ge fraction x, a critical discussion of the results' dependence on the used conduction- and valence-band parameters are beyond the scope of this work.

The dependence of electron and hole mobilities on the dopant concentrations and the temperature was included via the minimos-model provided by the nextnano++ package, using default parameters.[42] The resulting majority carrier mobility values for doping concentrations of $5\cdot10^{18}$ cm$^{-3}$ at T=200 K and 300K relevant for this work are listed in Tab. 2.

Table 2: Majority carrier mobility values for doping concentrations of $5\cdot10^{18}$ cm$^{-3}$ as used in this work. For the $Si_{1-x}Ge_x$ alloy, the corresponding values were obtained by linear interpolation with respect to x.

| Temperature [K] | majority carriers | mobility in Si [cm$^2$/Vs] | mobility in Ge [cm$^2$/Vs] |
|---|---|---|---|
| 200 | e | 181 | 1908 |
| | h | 100 | 385 |
| 300 | e | 162 | 1323 |
| | h | 85 | 329 |

For the electron-hole recombination, Shockley-Read-Hall (SRH)-, radiative- and Auger-processes were included in the simulations. Here, the parameters for the SRH recombination were adjusted so that the resulting SRH minority carrier lifetime values $\tau_n$ and $\tau_p$ (listed in Tab. 3) approximately agree with the experimentally observed minority carrier life times[43,44] in Si and Ge at doping concentrations of $5 \cdot 10^{18}$ cm$^{-3}$. We emphasize that in the experimental lifetime data, radiative- and Auger-recombinations contribute to some degree.[44] Thus, our parameter choice for SRH recombination results in an underestimation of carrier concentration and, thus, of radiative recombination rates.

From the self-consistent transport simulations, the position-dependence of the e- and h-currents and the e- and h- quasi-Fermienergies result for every voltage step. Based on these data, the spontaneous optical emission spectra were calculated.

Table 3: Electron - hole recombination parameters used in this work for doping concentrations of $5 \cdot 10^{18}$ cm$^{-3}$. For the $Si_{1-x}Ge_x$ alloy, the values were obtained by linear interpolation with respect to x.

| Recombination process | parameter | Si | Ge |
|---|---|---|---|
| Shockley-Read-Hall | $\tau_n$ [ns] | 223 | 60 |
| | $\tau_p$ [ns] | 36 | 56 |
| radiative | B [cm$^3$s$^{-1}$] | $4.7 \times 10^{-15}$ [45*] | $6.41 \times 10^{-14}$ [46] |
| Auger | $C_n$ [cm$^6$s$^{-1}$] | $3.41 \times 10^{-31}$ [47] | $1.1 \times 10^{-31}$ [48] |
| | $C_p$[cm$^6$s$^{-1}$] | $1.17 \times 10^{-31}$ [47] | $1.1 \times 10^{-31}$ [48] |

* 300 K-value from Ref. 45.

## Epitaxial growth

All DHS samples were grown in a Riber SIVA-45 solid source molecular beam (MBE) epitaxy system. The samples for atomic force microscopy surface investigation were grown on FZ Si(001) wafers with resistivities > 5000 Ωcm. The diode DHS structures wer grown on highly n-type (Si:As, $1 \cdot 10^{19}$ cm$^{-3}$) doped CZ Si(001), which provides the n-region of the p-i-n diode. After a substrate cleaning process, see Ref. 20, the substrates were loaded into the MBE chamber and degassed at 700 °C for 20 min. For samples for surface characterization, 10 nm of Si buffer was deposited at $T_G$, ramped from 450°C-550°C. For one AFM sample, we subsequently deposited 16 nm of $Si_{0.6}Ge_{0.4}$ at 350°C. We used Ge and Si rates of 0.15 Å/s and 0.225 Å/s, respectively that were calibrated using x-ray diffraction of calibration samples. For another AFM sample, we added an annealing step at 500°C in the middle of the ULT-grown $Si_{0.6}Ge_{0.4}$ layer, i.e., after 8 nm thickness. For the diode (sample C in the supplementraty material), we first grew 50 nm of Sb-doped Si buffer layer ($5\cdot10^{18}$ $cm^{-3}$), followed by 8 nm Sb-doped $Si_{0.6}Ge_{0.4}$ ($5\cdot10^{18}$ $cm^{-3}$), the aforementioned annealing step at 500°C and another 8 nm of B-doped $Si_{0.6}Ge_{0.4}$ ($5\cdot10^{18}$ $cm^{-3}$). The p-type region of the diodes was formed by 10 nm Si:B ($5\cdot10^{18}$ $cm^{-3}$), grown at 350°C, 40 nm Si:B ($5\cdot10^{18}$ $cm^{-3}$), grown at 350°C to 500°C and 150 nm of Si:B ($2\cdot10^{19}$ $cm^{-3}$), grown at 500°C. For the used doping concentrations, free carrier absorption of the near-infrared light emission is insignificant.

By lowering $T_G$ from 650°C to 350°C while maintaining pristine maximum chamber pressures $< 2\cdot10^{-10}$ mbar (base pressure; $5\cdot10^{-11}$ mbar) during growth, we recently increased the critical layer thickness from less than 2 nm to 16 nm for a $Si_{0.6}Ge_{0.4}$ alloy grown on Si.[20] However, despite their excellent morphological quality, the electronic properties of these layers determining their potential as active material in opto- and electronic devices have remained unexplored up to now. Therefore, as a proof-of-concept based on self-consistent Poisson-current bandstructure calculations, we here established for the first time a group-IV DHS light emitting diode that pushes the state-of-the-art with respect to high-temperature light emission towards technologically relevant temperatures above 300 K.

**Device fabrication and electrical characterization**

We processed the MBE-grown DHS material into square-shaped mesa (400×400 $\mu m^2$) devices for vertical current driving and used plasma deposition of 3 nm of Ti and 50 nm of Au as the top contact and also at the substrate's backside as the bottom contact, followed by In bonding to a chip carrier. No sidewall passivation was applied. Thus, Au wire bonding was performed to drive the device to the device's top. DHS LEDs were driven by a Keithley 236 source measure unit with a maximum current of 100 mA. Currents up to 0.1 A were applied to the mesa top contact in respect to the grounded bottom metallization. The voltage drop across the device was monitored along with the driving current, and measured as I-V curves. For I-V spectra and emission spectra, the DHS-LED was built into a cryostat to access a temperature range from 10 K to 360 K. The fused silica window induces an absorption line, visible at 1383 nm.

**Spectroscopic investigation**

The optical collection spot at the sample has a diameter of approximately 0.4 mm which corresponds to device dimensions. The EL signal is coupled into a multimode fiber, which, in turn, is connected to the ACTON SP300i spectrometer and the liquid nitrogen-cooled OMA V (Princeton Instruments) linear InGaAs photodiode array with a cut-off energy of 0.775 eV (1600 nm). For L-I curves, the diode was driven using the Keithley 236 sourcemeter, the signal coupled to a multi-mode fiber, modulated by a chopper and guided to a liquid nitrogen-cooled InAs point detector with a cutoff energy of 0.4 eV (~3 µm), i.e., far below the DHS emission band, see Fig. 4c,d.

**Transmission electron microscopy**

A ZEISS Crossbeam 1540XB scanning electron microscope with focused ion beam was used to prepare the {110} zone-axis TEM lamellae. The TEM experiments were carried out in a JEOL JEM-2200FS operated at an acceleration voltage of 200kV. The TEM is equipped with an in-column Ω-filter and a TVIPS TemCam-XF416 camera.

**Acknowledgments**

This research was funded in whole or in part by the Austrian Science Fund (FWF) [10.55776/Y1238] and [10.55776/P36608]. For open access purposes, the author has applied a CC BY public copyright license to any author-accepted manuscript version arising from this submission. The financial support by the Austrian Federal Ministry of Labour and Economy, the National Foundation for Research, Technology and Development and the Christian Doppler Research Association is gratefully acknowledged.

**Correspondence and requests for materials** should be addressed to Thomas Fromherz, Johannes Aberl, and Moritz Brehm.

**Supplementary Material**

# A group-IV double heterostructure light emitting diode for room temperature gain in Silicon

Andreas Salomon[1], Johannes Aberl[1], Lada Vukušić[1], Enrique Prado-Navarrete[1], Jacqueline Marböck[1], Diego-Haya Enriquez[1], Jeffrey Schuster[1], Kari Martinez[2], Heiko Groiss[2], Thomas Fromherz[1,*], and Moritz Brehm[1,*]

[1] *Institute of Semiconductor and Solid State Physics, Johannes Kepler University Linz, Altenberger Strasse 69, A-4040 Linz, Austria*

[2] *Christian Doppler Laboratory for Nanoscale Phase Transformations, Center for Surface and Nanoanalytics, Johannes Kepler University Linz, Altenberger Straße 69, 4040 Linz, Austria*

* Corresponding authors: *thomas.fromherz@jku.at ; moritz.brehm@jku.at*

**Structural characterization of the DHS structure.**

Figure S1 shows atomic force microscopy images related to the thermal stability of the Ge-rich DHS structures. Sample A on the left depicts the sample surface of a 16 nm thick $Si_{0.6}Ge_{0.4}$ layer, deposited at a growth temperature of 350°C. The root-mean square surface roughness (rms) is indicated in the micrographs and is, in all cases, < 0.14 nm, i.e., close to that of pristine Si wafers. Sample B in the middle also depicts the surface of a 16 nm thick $Si_{0.6}Ge_{0.4}$ layer grown on Si. However, for this layer, a temperature ramp to 500°C was performed after the growth of the first 8 nm of the SiGe alloy to mimic the thermal budget needed for the deposition of Ge quantum dots that can be grown at 500°C. We note that the surface roughness of the layer is entirely unaffected by this temperature step. Finally, sample C depicts an AFM image of the complete diode structure, i.e., after overgrowth of the DHS structure depicted in Fig. S1, sample B. All samples exhibit a completely flat sample surface and the absence of three-dimensional growth. The TEM investigation of the diode layer structure (see Fig. 4 of the main text) confirms that no dislocations are present in the DHS structure. The X-ray diffraction data of sample B are depicted as reciprocal space maps and a line scan in Fig. S2. The results clearly demonstrate the pseudomorphic nature of the layer, as evidenced by the narrow Qx-range of the SiGe-related features and the pronounced thickness oscillation fringes.

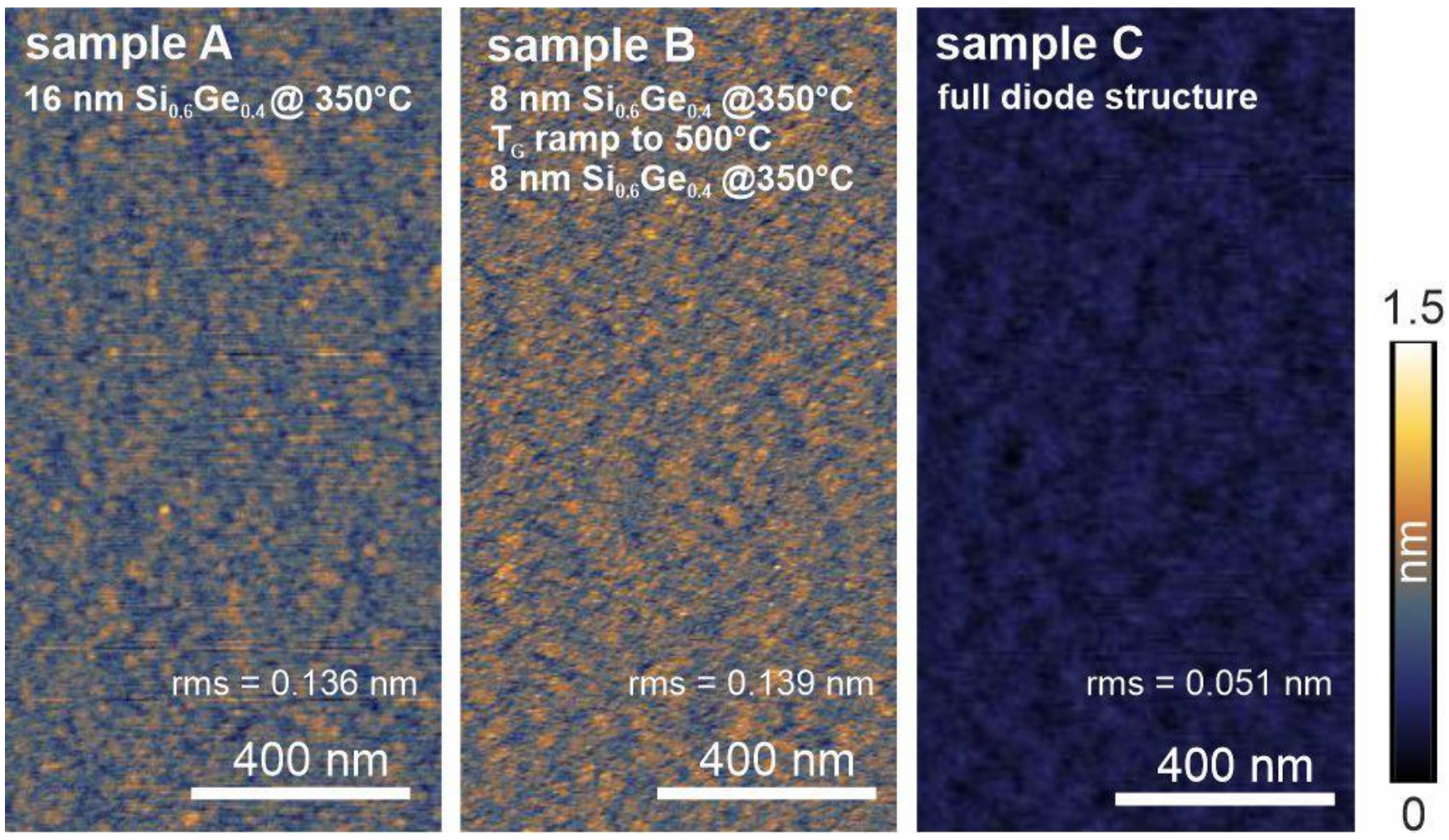

**Fig. S1| Thermal stability of the DHS structure.** AFM images of (sample A) a flat, 16 nm thick $Si_{0.6}Ge_{0.4}$ heterostructure grown at 350°C. Sample B: an added temperature ramp to 500°C in the middle of the 16 nm thick $Si_{0.6}Ge_{0.4}$ layer was performed. The SiGe layer grown at low temperatures sustains the thermal budget at 500°C. Sample C: AFM image of the surface of the doped and capped full diode structure, containing the sample sequence shown in sample B. In all three cases, the root mean square rms surface roughness is below 0.14 nm, i.e., close to that of a pristine Si(001) wafer.

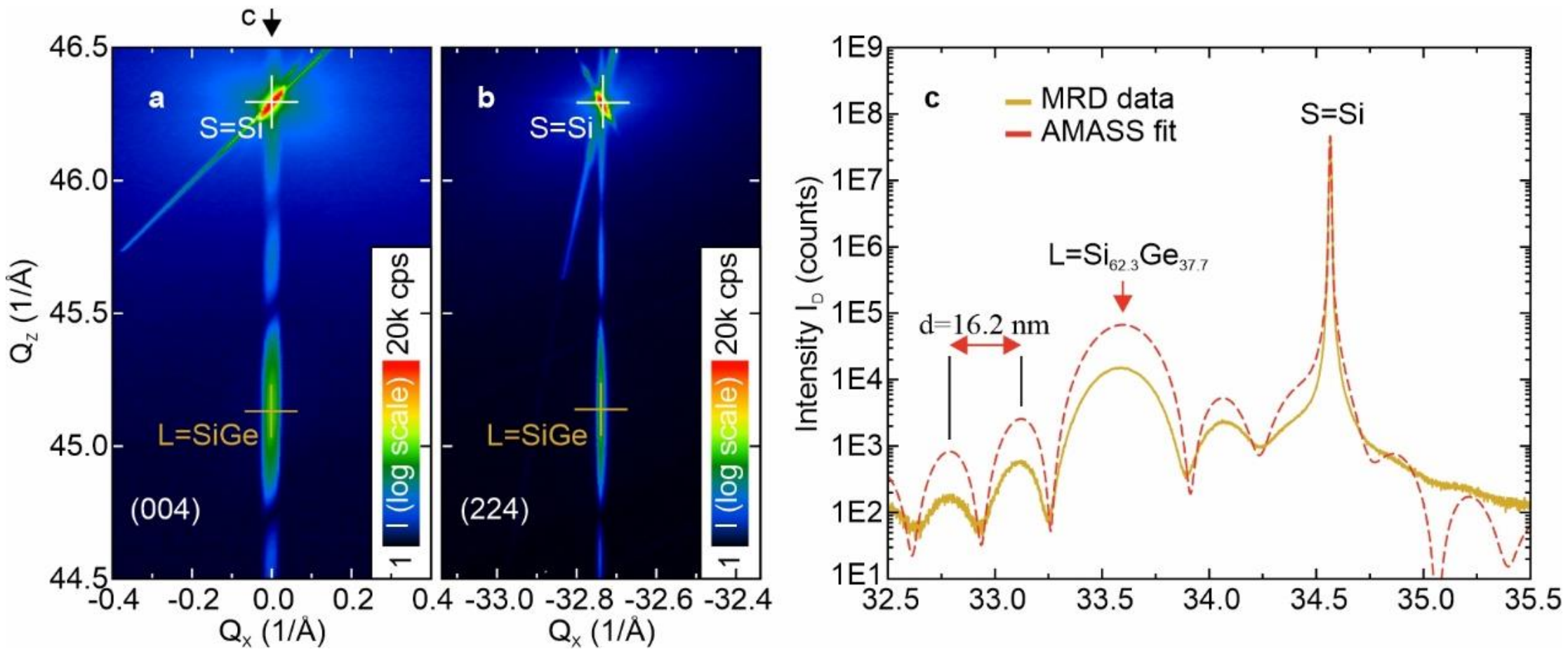

**Fig. S2| Pseudomorphic nature of the strained SiGe/Si layers.** Structural properties of a 16 nm thick $Si_{0.6}Ge_{0.4}$ film. **a,b** X-ray diffraction (XRD) reciprocal space maps in (224) and (004). Fully pseudomorphic growth is evidenced by the perfect agreement of the $Q_x$ components of the Si and SiGe X-ray scattering vectors in both maps, as well as by the small width of the SiGe scattering signal in $Q_x$ direction. Bright, straight lines in the reciprocal space maps are artifacts from the experimental setup. **c** XRD line scan along the (004) direction showing the pseudomorphic SiGe-related peak and the thickness oscillation fringes. The direction of the line scan is indicated by the black arrow on top of panel **a**.

**Dot-in-well heterostructures**

For III-V QD lasers, the best light-emission intensities are typically obtained for type-I dot-in-well heterostructures, something considered not accessible in the SiGe system. However, Figs. S3, S4 show auspicious signs that at least small S-K nanostructures like conventional hut-clusters can be readily embedded into type-I DHS. First, we tested that dense hut-cluster carpets can be grown (at $T_G$ = 500°C) on an 8 nm thick ULT-grown $Si_{0.6}Ge_{0.4}$ layer, i.e., into the middle of the DHS (see Fig. S3 middle panel). Interestingly, compared to the growth on Si, slightly less Ge is needed to obtain full areal coverage of the hut clusters (before the nucleation of dislocated QDs). This finding indicates that (i) the strain is preserved in the lower cladding layer and (ii) the surface energies and chemical potential play a role in the growth of Ge QDs on SiGe surfaces. Figure S3 on the right side depicts the Ge QD layer after subsequent overgrowth with another 8 nm of $Si_{0.6}Ge_{0.4}$, deposited at $T_G$ = 350°C. This result confirms that conformal overgrowth is possible and that the thermal budget of the DHS can sustain temperature steps at least ~500°C.

Figure S4 shows cross-sectional transmission electron microscopy images of the fully grown dot-in-DHS device. One can clearly see that the lower 8 nm and the upper 8 nm of the SiGe structures are separated by Ge-rich (i.e., darker) regions of ~1-2 nm height, i.e., small Ge QDs. The total dot-in-well structure has, due to the presence of the QDs, a slightly larger thickness than the 16 nm of the DHS structure, described in the main part of this work. Both TEM images also demonstrate the fully dislocation-free growth of this dot-in DHS structure.

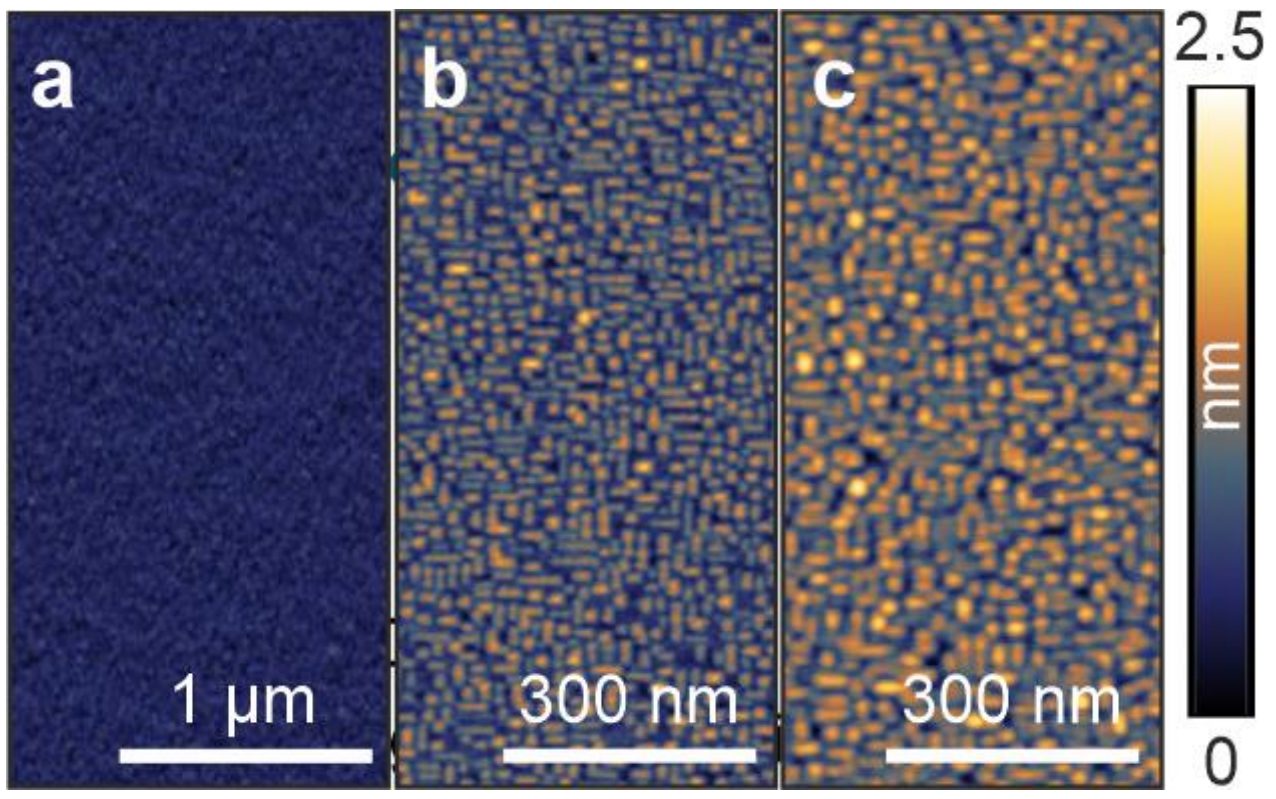

**Fig. S3| Growth and overgrowth of dot-in-DHS structure.** AFM images of **a** 16 nm SiGe40% at $T_G$ = 350°C. After 8 nm of SiGe growth, $T_G$ was raised for 2 min to 500°C to simulate active emitter growth. **b** Growth of Ge QDs (hut clusters grown at $T_G$ = 500°C) onto of a layer of 8 nm of SiGe40%. **c** Conformal overgrowth of the Ge QDs in by another 8 nm of SiGe 40%, grown at $T_G$ = 350°C.

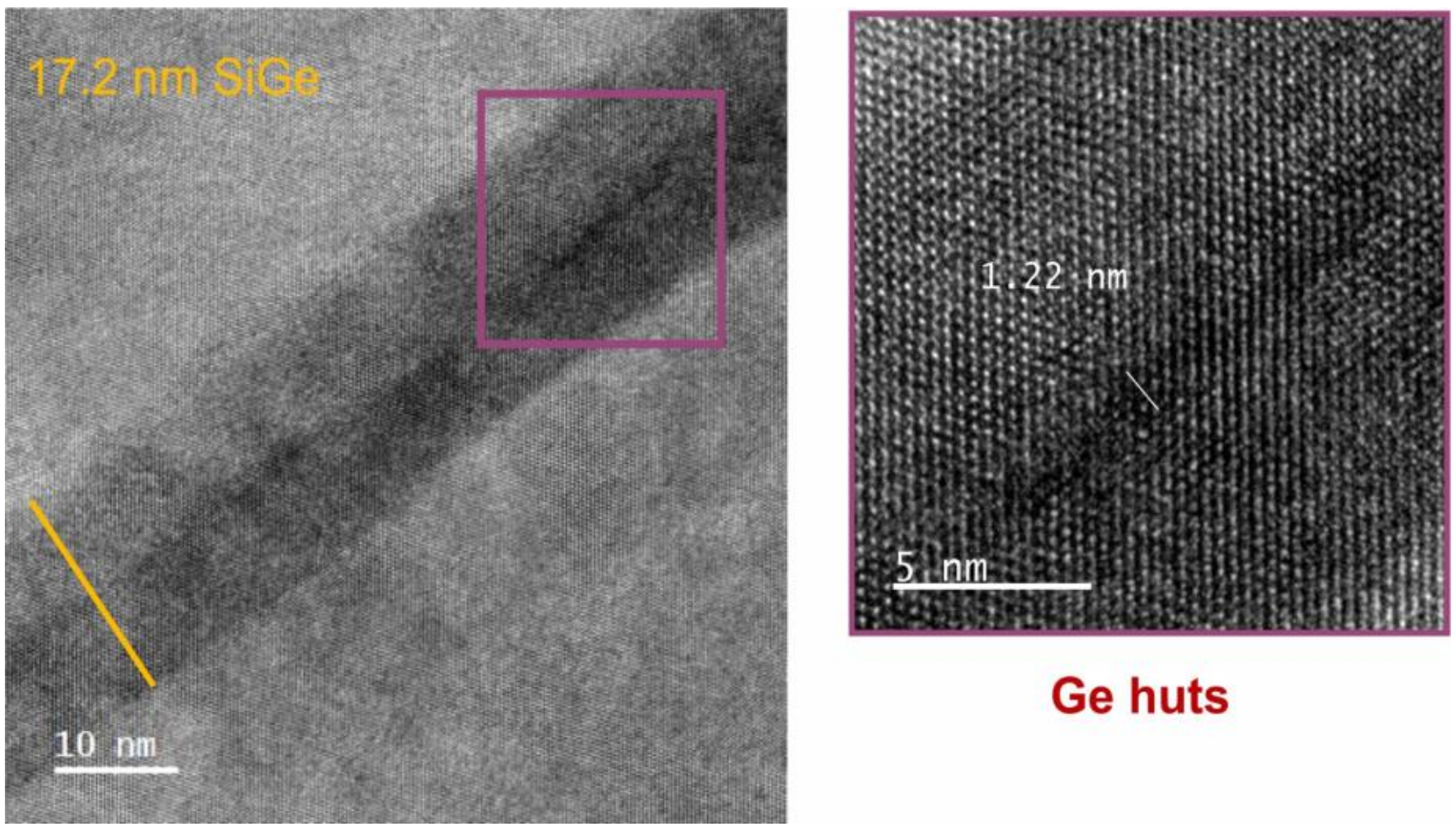

**Fig. S4| Transmission electron microscopy characterization of the dot-in-DHS structure.** Cross-sectional conventional and high-resolution TEM images of a 16 nm thick $Si_{0.6}Ge_{0.4}$ layer into which QDs were incorporated. The layer growth sequence comprises of a Sb-doped Si buffer layer, 8 nm of SiGe40% at $T_G$ = 350°C, 0.6 nm of Ge at $T_G$ = 500°C, 8 nm of SiGe40%, and a B-doped Si capping layer.

After device formation, we measured the optoelectronic properties of this group-IV dot-in-DHS diode structure, i.e., I-V-, light-current (L-I) curves, and EL spectra (Figs. S5a,b). For different sample temperatures ranging from 10 K to 360 K (87°C), the left ordinate in Fig. S5a depicts the bias applied to the device, i.e., the V-I curve, and the right ordinate the EL emission intensity, i.e. the L-I curve. Figure S5b depicts the EL emission spectra recorded at the respective sample temperature. We note that EL spectra are cut off due to the bandgap of the InGaAs line detector, while the L-I curves in Fig, S5a were measured with an InAs point detector, covering the whole emission range. Based on the emission spectra and the integrated EL intensities shown in Fig. S5a, we highlight that thermal quenching of the EL is absent in these dots-in-DHS devices, at least up to 360 K, indicating that minority carrier injection can be efficiently suppressed in this device structure. A scheme of the sample layout is shown in the inset in Fig. S5b.

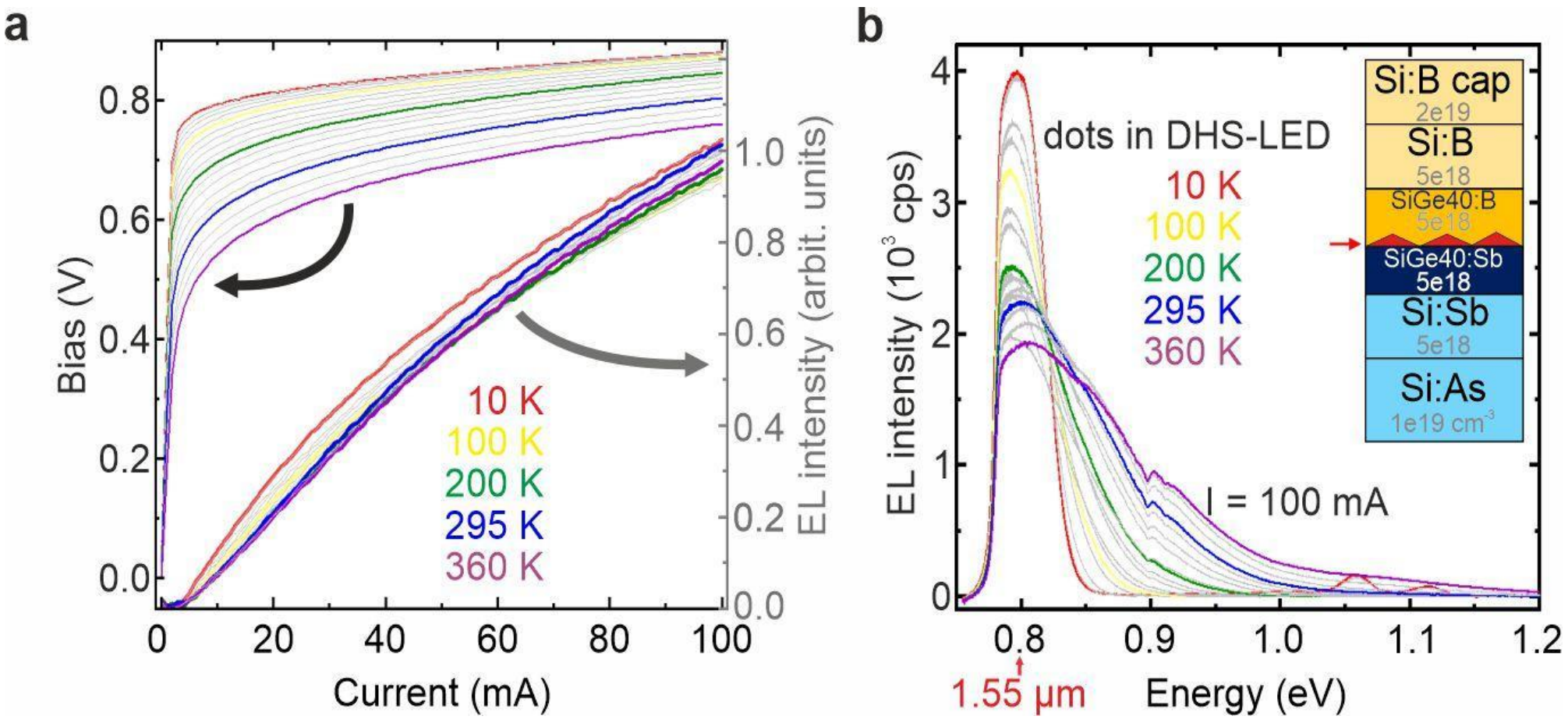

**Fig. S5| Electrical and optoelectronic characterization of the dot-in-DHS structure. a** Current-voltage characteristics, i.e., I-V curves and light-current (L-I) curves. The L-I curves were measured with an InAs point detector, covering the entire emitter emission range. **b** Electroluminescence spectra of dot-in-well heterostructure. The emission occurs right at the telecom C-band at a wavelength of 1550 nm. The cut-off of the InGaAs line detector is at 0.77 eV. Thermal quenching of the EL is absent in these dot-in-DHS devices, at least up to 360 K, i.e., the limit of the setup.

## Increasing the Ge content in the DHS layer

The simulations in the main text predict that the achievable optical gain in a DHS embedded in a Si diode increases with the Ge content of the SiGe DHS layer. Higher Ge concentrations, however, are associated with higher strain in the SiGe layer; thus, already at lower layer thicknesses, the layers are expected to relax. The theoretical prediction of the exponential relationship between Ge content and layer thickness before relaxation was described for the growth of SiGe alloys on Si by Matthews and Blakeslee [S1] and People and Bean [S2,S3]. However, until recently, these curves were only experimentally verified only for SiGe alloys with very low Ge contents <40%. For higher Ge concentrations, elastic relaxation, i.e., QD formation, becomes a dominant relaxation mechanism, especially at typically used growth temperatures of ≥500°C [20]. For example, QD formation occurs already at a thickness of ~1.8 nm for a SiGe alloy with 40% Ge content grown on Si(001). For Ge on Si(001), the transition from two-dimensional (2D) to three-dimensional (3D) growth occurs already at ~0.5 nm.

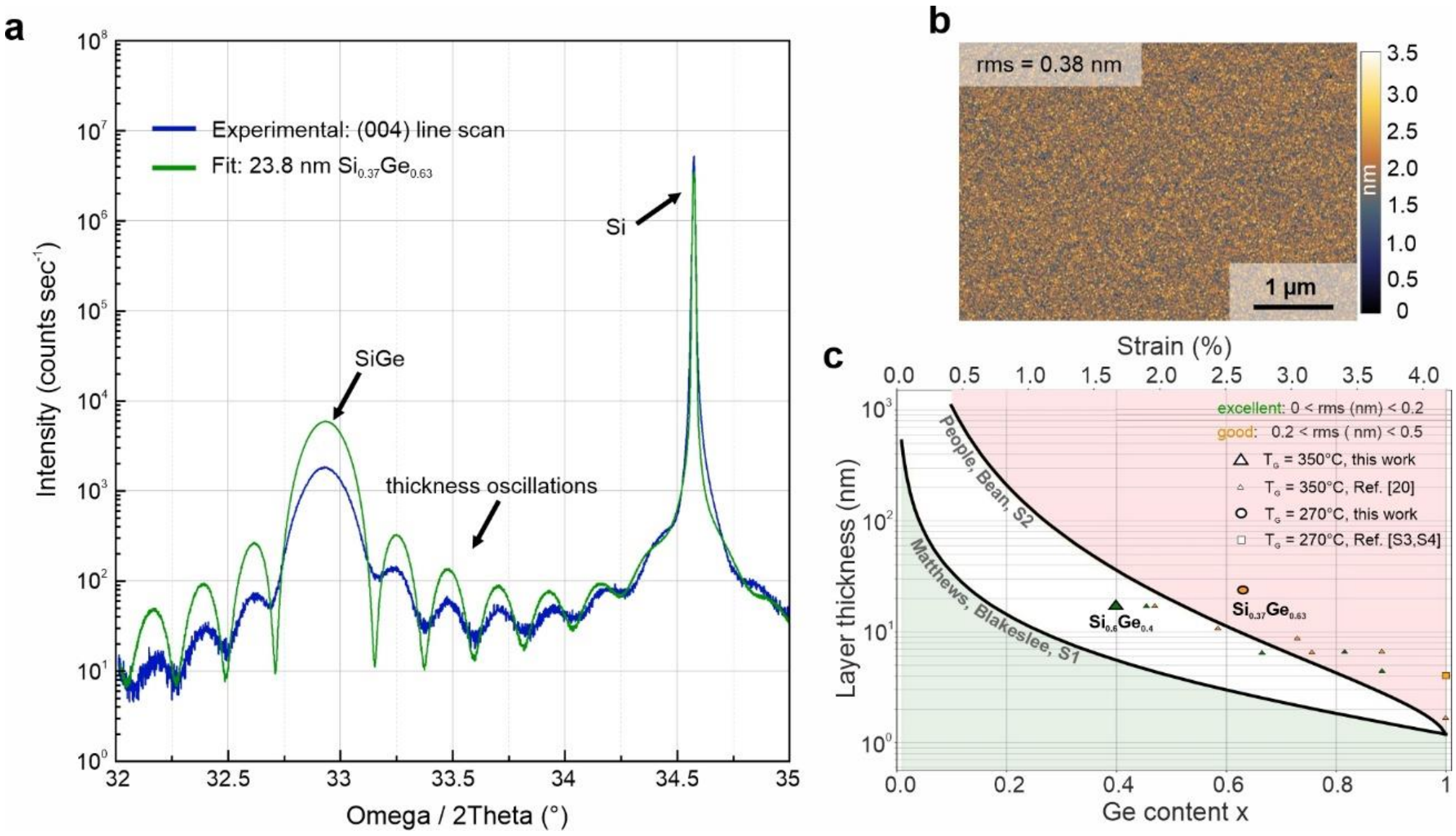

**Fig. S6| Thick, pseudomorphic SiGe layers with high Ge content. a** X-ray diffraction (XRD) (004) line scan of a 23.8 nm thick SiGe layer with ~63% Ge content grown on Si(001) at $T_G$ = 270°C. The line scan shows the pseudomorphic SiGe-related peak and thickness oscillation fringes. **b** AFM of the sample surface of this 23.8 nm thick SiGe layer with 63% Ge content, indicating a low surface roughness of <0.4 nm. **c** Critical layer thickness versus Ge content x of epitaxial Ge-rich $Si_{1-x}Ge_x$ layers grown on Si(001). The black lines indicate theoretical results from Matthews–Blakeslee [S1] and People–Bean [S2,S3]. Large data points (green triangle at x = 0.4 and orange circle at x = 0.63) correspond to the samples investigated in the main text and the supplementary material. Small triangles were obtained in Ref. [20] for $T_G$ = 350°C. Lowering the $T_G$ further allows for an increased layer thickness; see data points for 270°C at x = 0.63 and x = 1.

These 2D layer thicknesses are far too low for useful DHS structures, as strong quantum confinement would nullify the benefit of the potential barrier that limits minority carrier injection (see Fig. 5b). For useful DHS, SiGe alloy layers with excellent crystalline quality and with at least 10-15 nm thickness are needed.

As the main text and Ref. [20] discussed, ultra-low growth temperatures allow for extensive strained layer supersaturation. In Figure S6, we discuss the structural properties of a <20 nm thick SiGe layer with a Ge content of 63%. We note that such a Ge content corresponds to the red curves/points in Fig. 3b of the main text and a resulting room temperature gain of ~300 $cm^{-1}$. Figure S6(a) shows an XRD (004) line scan of a ~24 nm thick $Si_{0.37}Ge_{0.63}$ layer, deposited at $T_G$ = 270°C on a Si(001) buffer layer. The layer is pseudomorphic, as evidenced by the observed thickness oscillation fringes that are in excellent agreement with the simulated ones. The AFM image of the sample surface of the strained $Si_{0.37}Ge_{0.63}$ layer shows a slightly increased but still excellent surface roughness of 0.38 nm. The slight increase of the rms with respect to the results for the layer with x = 0.4, see Fig. S1, could indicate that the layer is at the onset of elastic relaxation. However, the XRD results in Fig. S6(a) show that the layer is still pseudomorphically strained, and the AFM image in Fig. S6(b) clearly shows the absence of misfit dislocations that would be visible as a cross-hatch at the sample surface. Pure Ge layers of 4 nm thickness and similar surface roughness (see square data point in Fig. S6(c)) were used to fabricate excellent nanoelectronics devices based on reconfigurable transistors [S3,S4]. Thus, the results in Fig. S6 demonstrate the excellent epitaxial quality of the Ge-rich thin films, grown at temperatures as low as 270°C and at excellent growth pressures; see Fig. S7 for the 23 nm thick $Si_{0.37}Ge_{0.63}$ layer grown on Si.

Figure S6(c) shows the layer thickness versus Ge content x of epitaxial Ge-rich $Si_{1-x}Ge_x$ layers grown on Si(001). The black lines indicate theoretical results from Matthews–Blakeslee [S1] and People–Bean [S2,S3]. Large data points (green triangle at x = 0.4 and orange circle at x = 0.63) correspond to the samples investigated in the main text and the supplementary material. Small triangles were obtained in Ref. [20] for $T_G$ = 350°C. Lowering the $T_G$ further results in an increased layer thickness; see data points for 270°C at x = 0.63 and x = 1.

Careful future investigations of the critical thickness of SiGe/Si(001) layers depending on parameters like growth temperature, growth rate, and growth pressure will likely lead to a further expansion of the frontiers of fully strained 2D SiGe epitaxy on Si and their application in nanoelectronics and optoelectronics.

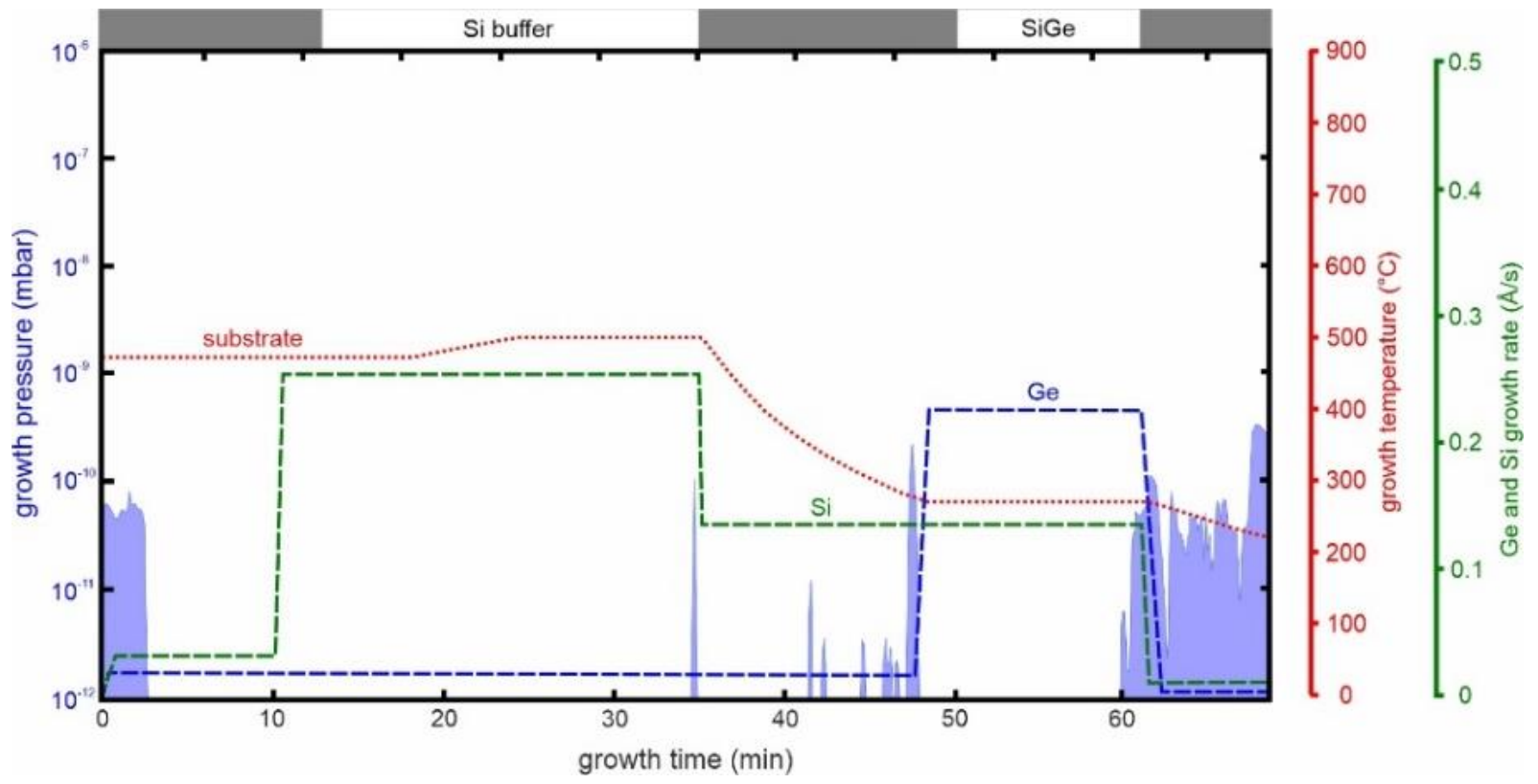

**Fig. S7| Growth protocol of the 23 nm thick $Si_{0.37}Ge_{0.63}$ layer grown on Si.** Low growth pressures enable excellent layer qualities even for ultra-low-temperature growth. The blue-shaded areas indicate the growth pressure (left ordinate). The right ordinates correlate to the substrate temperature (red dotted line), and the Si and Ge growth rate (green and blue dashed lines). Within the grey-shaded times, $T_G$ or the emission rates were ramped to their respective setpoints with evaporators closed.